# Experimental evidence of monolayer arsenene: An exotic two-dimensional semiconducting material

J. Shah*, W. Wang*, H.M. Sohail and R.I.G. Uhrberg, Department of Physics, Chemistry, and Biology, Linköping University, S-581 83 Linköping, Sweden

**Abstract:** Group V element analogues of graphene have attracted a lot attention recently due to their semiconducting band structures, which make them promising for next generation electronic and optoelectronic devices based on two-dimensional materials. Theoretical investigations predict high electron mobility, large band gaps, band gap tuning by strain, formation of topological phases, quantum spin Hall effect at room temperature, and superconductivity amongst others. Here, we report a successful formation of freestanding like monolayer arsenene on Ag(111). This was concluded from our experimental atomic and electronic structure data by comparing to results of our theoretical calculations. Arsenene forms a buckled honeycomb layer on Ag(111) with a lattice constant of 3.6 Å showing an indirect band gap of ~1.4 eV as deduced from the position of the Fermi level pinning.

The isolation of two-dimensional (2D) carbon in the form of a single layer honeycomb structure (graphene)[1] was the starting point of today's intensive research on various 2D materials. In particular, the electronic properties, i.e., the high conductivity in combination with the linear dispersion of the π-band, forming a Dirac cone, are highly interesting for applications in nanoelectronics[2]. However, this utilization of graphene is severely hampered by the lack of an intrinsic band gap, which it shares with graphene like structures of other group IV atoms, i.e., Si, Ge, and Sn[3].

Corresponding authors:
Jalil Shah: jalil.shah@liu.se
Weimin Wang: weimin.wang@liu.se

Layered materials formed by the group V atoms P, As, Sb and Bi, offer a possible solution to this problem. Recently, few-layer phosphorus sheets were mechanically exfoliated from black phosphorus, which is a layered material held together by van der Waals (vdW) forces[4]. Each layer has a puckered structure in contrast to the honeycomb structure of graphene. Theoretically, single layer phosphorus (phosphorene) is predicted to have a direct band gap in the range 0.8 to 2.35 eV [4-8] depending on calculation method. Experimental values for the optical and quasiparticle band gaps of 1.3 and 2.2 eV, respectively, were reported in Ref. 9, while a band gap value of 2 eV has been reported from scanning tunneling spectroscopy (STS) measurements[6]. Experimental studies have been performed both on few-layer[9-11] and on monolayer phosphorene[4,9]. However, the practical use of phosphorene in nanoelectronics might be difficult due to the methods of producing few-layer and monolayer phosphorene. Exfoliation from black phosphorus is the prevailing procedure, which results in small flakes of 1-10 μm size[4,9,12]. Experimental investigations have also been performed on antimonene (Sb)[13-15] and bismuthene (Bi)[16]. Extended single layers of antimony and bismuth grown on Ag(111) and SiC(0001), respectively[15,16], have been interpreted as planar honeycomb structures with significantly larger lattice constants compared to the theoretically predicted buckled structures[7].

There is not yet any experimental report of a single layer honeycomb structure formed by As, i.e., arsenene. Theoretically, single layer arsenene prefers a buckled honeycomb structure[7], while puckered[18-21] and planar[18] arsenene are less favorable energetically. The values of the lattice constant ($a_{ars}$) and the buckling, d, depend on the theoretical method. Values for $a_{ars}$ range from 3.54 to 3.64 Å[17-19,21-26], and d ranges from 1.35 to 1.4 Å[17,21,23-27]. Arsenene has an indirect gap with calculated values between 1.42 and 1.88 eV when using the Perdew-Burke-Ernzerhof (PBE) functional [18,19,21,22,24-27] and between 2.0 and 2.49 eV when Heyd-Scurseria-Ernzerhof hybrid functionals (HSE/HSE06) were employed [17,19,25]. Strain effects on the electronic structure have been discussed in several papers[17-19,,23,25,28]. Apart from a possible conversion to a direct band gap semiconductor, strain has also been studied as a means to transform arsenene to a topological insulator (TI) and to achieve a quantum spin Hall (QSH) phase[23,27,28]. Furthermore, superconductivity with a transition temperature as high as 30.8 K has been predicted for strained electron doped arsenene[22]. Other theoretical studies deal with defects in the arsenene layer[21,29], and the formation of electrical contacts by investigating arsenene metal interfaces[24]. Furthermore, in a theoretical study by Pizzi *et al*, arsenene was identified as a promising 2D material for field effect transistors (FET) with application in future nano electronics[26].

In this paper, we present evidences of the successful formation of monolayer arsenene on Ag(111). This conclusion is derived from low energy electron diffraction (LEED), STM and ARPES in combination with density functional theory (DFT) calculations of the atomic and electronic structures. LEED data, shown in Fig. 1, provide evidence for the formation of a well-ordered arsenic structure. In addition to the 1×1 spots from Ag(111), shown in Fig. 1a, several diffraction spots of varying intensity appear after exposure to arsenic. In particular, there are six bright spots, of which one is indicated by a green arrow in Fig. 1b. It is natural to assign these bright spots to diffraction from a 1×1 unit cell of the arsenic structure. The diffraction pattern in Fig. 1b can be described as a 4×4 periodicity with respect to the ordered arsenic layer. It is worth noting that some of the 4× spots coincide with 1×1 spots of Ag(111), compare the red arrows in Figs. 1a and 1b. From the distances between Ag(111) 1×1 spots in Fig. 1a and distances between arsenic related 1×1 spots in Fig 1b, we find that the ratio between the surface lattice constant of Ag(111) ($a_{Ag111}$=2.89 Å) and that of the arsenic structure ($a_{ars}$) is very close to 0.8. Based on these observations, we can conclude that the arsenic structure has a lattice constant that is $a_{ars}$≈1.25×2.89 Å≈3.61 Å. This experimental value compares very well with theoretical values for the lattice constant of freestanding arsenene, ranging from 3.54 to 3.64 Å [17-19,21,22,24]. LEED data clearly indicate that an arsenene layer has formed with the same orientation of the unit cell as that of the Ag(111) substrate, with a "perfect" match of $4a_{ars}$ to $5a_{Ag111}$. This difference in the lattice constants is the origin of the Moiré type of 4×4 LEED pattern that is observed for arsenene on Ag(111). In the following, we present STM and ARPES data, which, in combination with theoretical results, further verify the formation of arsenene.

Arsenene was found to grow uniformly across the Ag(111) surface. A typical STM image is shown in Fig. 2a. This overview shows the arsenene layer on two terraces of the Ag(111) substrate. The resolution is not sufficient to show the atomic structure, but some typical defects are visible. There are a few dark triangular defects corresponding to missing arsenic atoms. A white arrow in Fig. 2a indicates one such defect. Another type of defect, indicated by a blue arrow, appears as dark lines. These lines have mainly three different orientations, consistent with a hexagonal structure. The zoom-in image in Fig. 2b reveals the structure responsible for the 4× periodicity observed by LEED. There is a hexagonal arrangement of rather big bright features separated by a distance corresponding to 4× $a_{ars}$. Fast Fourier transforms (FFTs) of STM images such as the one in Fig. 2b show Fourier components that generate a pattern closely matching the diffraction pattern obtained by LEED, c.f., Fig. 1b and Fig. 2c. The line defects in Fig. 2b separate ordered domains of the 4× structure. There appears to be a small phase shift in the positions of the bright features creating the linear boundaries. However, a detailed structure cannot be derived from Fig. 2b. The appearance of the 4× structure is sensitive to the bias used to obtain the STM data. When reducing the bias, the 4×4 structure fades out and a 1×1 arsenene structure becomes clear, see the inset of Fig. 2b. This image shows a hexagonal structure consistent with buckled arsenene. The apparent heights of the arsenic atoms are modulated by a 4×4 Moiré periodicity originating from the difference in the lattice constants of Ag(111) and arsenene. An experimental lattice constant of arsenene can be deduced from the separation between the atomic features that are well resolved in Fig. 2d. From the topographic profile (Fig. 2e), obtained along the black line in Fig. 2d, we find an average lattice constant of 3.6 Å, which agrees with the value derived from LEED.

As a final step to verify the formation of arsenene on Ag(111) we compare band structure data from ARPES with the calculated band structure of buckled arsenene. ARPES data were obtained along the high symmetry lines of the surface Brillouin zone (SBZ) corresponding to arsenene. Figs. 3a and 3b, show data and SBZs with symmetry lines, respectively. The ARPES data show the well-known contribution from the Ag substrate, labelled B, originating from direct transitions between Ag bulk bands. All other features correspond to emission from the arsenene. It is of major importance to note the dispersive band centered around $k_{//}=0$ Å$^{-1}$, labelled $V_1$, and that it reappears centered around $k_{//} = 2.0$ Å$^{-1}$, which corresponds to the distance in k-space between $\bar{\Gamma}_1$ and $\bar{\Gamma}_2$. From this experimental value of the reciprocal lattice dimension, we derive a real space lattice constant of 3.6 Å in agreement with the LEED and STM results. Apart from $V_1$, there is emission forming a broad dispersive band labelled $V_2$. The dispersion follows the periodicity of the arsenene SBZ as verified by the symmetry around the $\bar{M}$-point. Furthermore, a weak but noticeable intensity, C, appears at the Fermi level in the outer part of the SBZ along $\bar{\Gamma} \to \bar{M}$, but not along $\bar{\Gamma} \to \bar{K}$.

For comparison, we have calculated the band structure of arsenene from the model in Fig. 4a. Using the lattice constant derived from the LEED data, 3.61 Å, resulted in a buckling of 1.45 Å after full relaxation. This value is similar to calculated values in the literature, 1.35 Å[17], 1.38 Å[24] and 1.4 Å[21,25]. Figure 4b shows the band structure along $\bar{K}\bar{\Gamma}\bar{M}$ of the relaxed arsenene structure. As revealed by this band structure, arsenene is an indirect band gap semiconductor. Our calculation results in an indirect band gap of 1.47 eV, which falls within the range of the major part of the published values (1.42 – 1.71 eV[18,19,21,22,24,25]) from calculations using the PBE functional. There are three valence bands in the energy region covered by the experimental data.

Two of the bands, $A_1$ and $A_2$, have their energy maxima at $\bar{\Gamma}$, while the third band, $A_3$, has a local minimum at $\bar{\Gamma}$.

Figure 4c shows a comparison between the experimental dispersions and the calculated arsenene bands along the two high symmetry lines, $\bar{\Gamma}\bar{M}$ and $\bar{\Gamma}\bar{K}$. Close to $\bar{\Gamma}_1$ the intensity of $V_1$ fades out and it is difficult to determine the shape of the band. This band is more intense near $\bar{\Gamma}_2$, see Fig. 3a, where the shape is clearly visible and found to agree closely with that of the steepest calculated band $A_1$. The broad band $V_2$ is well reproduced by the less dispersive $A_3$ band, having a slightly lower energy at $\bar{K}$ compared to $\bar{M}$. The orbital content of the individual bands was investigated in detail in Ref. 25. Along $\bar{\Gamma}\bar{M}$, the $A_1$ band is composed of $p_x$ orbitals, $A_2$ of $p_y$ orbitals, and the $A_3$ band is derived from $p_z$ orbitals, as indicated in Fig. 4b. The geometry of the ARPES experiment is such that the A-vector of the linearly polarized light lies within the emission plane defined by the surface normal and the direction toward the electron analyzer. For the $\bar{\Gamma}\bar{M}$ data, the emission plane corresponds to a mirror plane of the arsenene structure. Using dipole selection rules it is possible to predict photoemission cross sections. An orthogonal projection of the A-vector results in a component perpendicular to the surface, $A_{perp}$, which excites states of $p_z$ character, i.e., the $A_3$ band, and in a component parallel to the surface, $A_{para}$, which excites p-states that lie within the emission plane, i.e., $A_1$ derived from $p_x$ orbitals and identified as the experimental $V_1$ band. The $A_2$ band derived from $p_y$ orbitals, having a node at the emission plane, is not excited which explains why this band is not observed in the experiment. The ARPES data is recorded by a fixed analyzer which implies that the incidence angle of the synchrotron light differs for data obtained along the $k_{//}$ axis resulting in changes of the projections of the A-vector. $A_{perp}$ decreases for increasing values of $k_{//}$ to become zero at $\approx 1.7$ Å$^{-1}$, which explains why the $V_2$ emission fades out as this value is approached.

Since the band structure of arsenene is semiconducting, the Fermi level of the arsenene/Ag(111) system could in principle be positioned anywhere within the band gap. However, the energies of the experimental bands are ~1 eV lower than the calculated values, which indicates a Fermi level pinning at or close to the conduction band minimum of arsenene. The position of the calculated band structure relative to the ARPES data in Fig. 4c, has been chosen to give a good match of the energies at the $\overline{K}$ and $\overline{M}$ points and, at the same time, resulting in a good match of the dispersion of $V_1$ and the corresponding calculated band, $A_1$. It is here interesting to note that the calculated conduction minimum is close to the C emission both in energy and $k_{//}$. This strongly suggests that the Fermi level of the arsenene/Ag(111) system is actually pinned at the conduction band minimum of arsenene. The fact that the theoretical conduction band minimum is located higher than the experimental band minimum, C, indicates that the experimental bandgap is less than the calculated one by ~0.1 eV.

In summary, the atomic and electronic structures of the arsenic layer on Ag(111) were experimentally investigated by LEED, STM and ARPES. We found a highly ordered 2D material showing semiconducting band structure. By comparing to results of theoretical calculations of arsenene, employing first principles DFT+PBE, we conclude that freestanding like monolayer arsenene has been successfully formed on Ag(111). The Fermi level of the arsenene/Ag(111) system was found to be pinned at the conduction band minimum and an experimental value of the indirect band gap of ~1.4 eV was deduced. Our report on arsenene formation opens up for extensive explorations of this promising material for next generation electronics and optoelectronic devices. Theoretical investigations in the literature predict high electron mobility, large band gap, band gap tuning by strain, formation of topological phases, quantum spin Hall effect at room temperature and superconductivity amongst other properties. Hence, we expect that our study will be of significant interest because of its potentially large impact on future nanoelectronics based on 2D materials.

**Methods**

**Experiments.** Samples were prepared *in-situ* in ultrahigh vacuum (UHV) systems with base pressures in the $10^{-11}$ Torr range. The Ag(111) crystal was cleaned by repeated cycles of sputtering by Ar+ ions (1 keV) and annealing at approximately 400 °C until a sharp (1×1) LEED pattern was obtained. Arsenene was formed on the Ag(111) substrate by exposure to an arsenic pressure of $2\times10^{-7}$ Torr, while keeping the substrate at 250-350 °C during 3 minutes. STM images were recorded at room temperature using an Omicron variable temperature STM at Linköping University, Sweden. ARPES data were obtained at the MAX-lab synchrotron radiation facility in Lund, Sweden, using the beam line I4 end station. Data were acquired at room temperature by a Phoibos 100 analyzer from Specs with a two-dimensional detector mounted at a 50° angle with respect to the incoming synchrotron light. The energy and angular resolutions were 50 meV and 0.3°, respectively.

**Calculations.** DFT calculations were performed to investigate the band structure of freestanding monolayer arsenene. The structure was modeled by a periodic slab with an in plane lattice constant of 3.61 Å and 19 Å of vacuum spacing. All atoms were relaxed until the average force was within 0.01 eV/Å. The band structure was calculated using the functional of Perdew, Burke and Ernzerhof (PBE) and the projector augmented wave method Vienna ab initio simulation package code (VASP)[30]. The energy cutoff of the plane-wave basis set was 434 eV, and the k-point mesh was (9×9×1).

**Acknowledgements**

Technical support from Dr. Johan Adell, Dr. Craig Polley and Dr. T. Balasubramanian at MAX-lab is gratefully acknowledged. Financial support was provided by the Swedish Research Council (Contract No. 621-2014-4764) and by the Linköping Linnaeus Initiative for Novel Functional Materials supported by the Swedish Research Council (Contract No. 2008-6582). The calculations were carried out at the National Supercomputer Centre (NSC), supported by the Swedish National Infrastructure for Computing (SNIC).

**Author contributions**

J. S. conducted the experiments in collaboration with H.M.S., W.W., and R.I.G.U. W.W. did the theoretical calculations with input from the other authors. All authors took part in data analysis and writing of the paper. R.I.G.U. is the group leader.

**Additional information**

Correspondence and requests for experimental work should be addressed to J. S.

Correspondence and requests for theoretical study should be addressed to W. W.

**Competing financial interests**

The authors declare no competing financial interests.

**Figures**

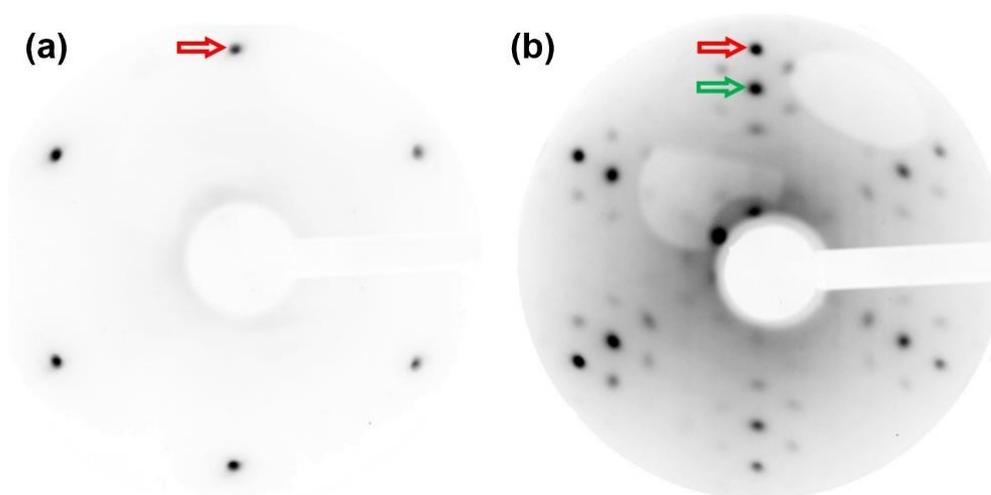

**Figure 1 | Low energy electron diffraction. a,** LEED pattern of clean Ag(111) at 50 eV showing 1×1 spots. **b,** LEED pattern of arsenene on Ag(111) at 50 eV. The position of one of the Ag(111) 1×1 spots is indicated by the red arrow while an arsenene 1×1 spot is indicated by the green arrow. The LEED pattern corresponds to a 4×4 periodicity with respect to the arsenene 1×1 structure. Ag(111) 1×1 spots overlap with 4× spots.

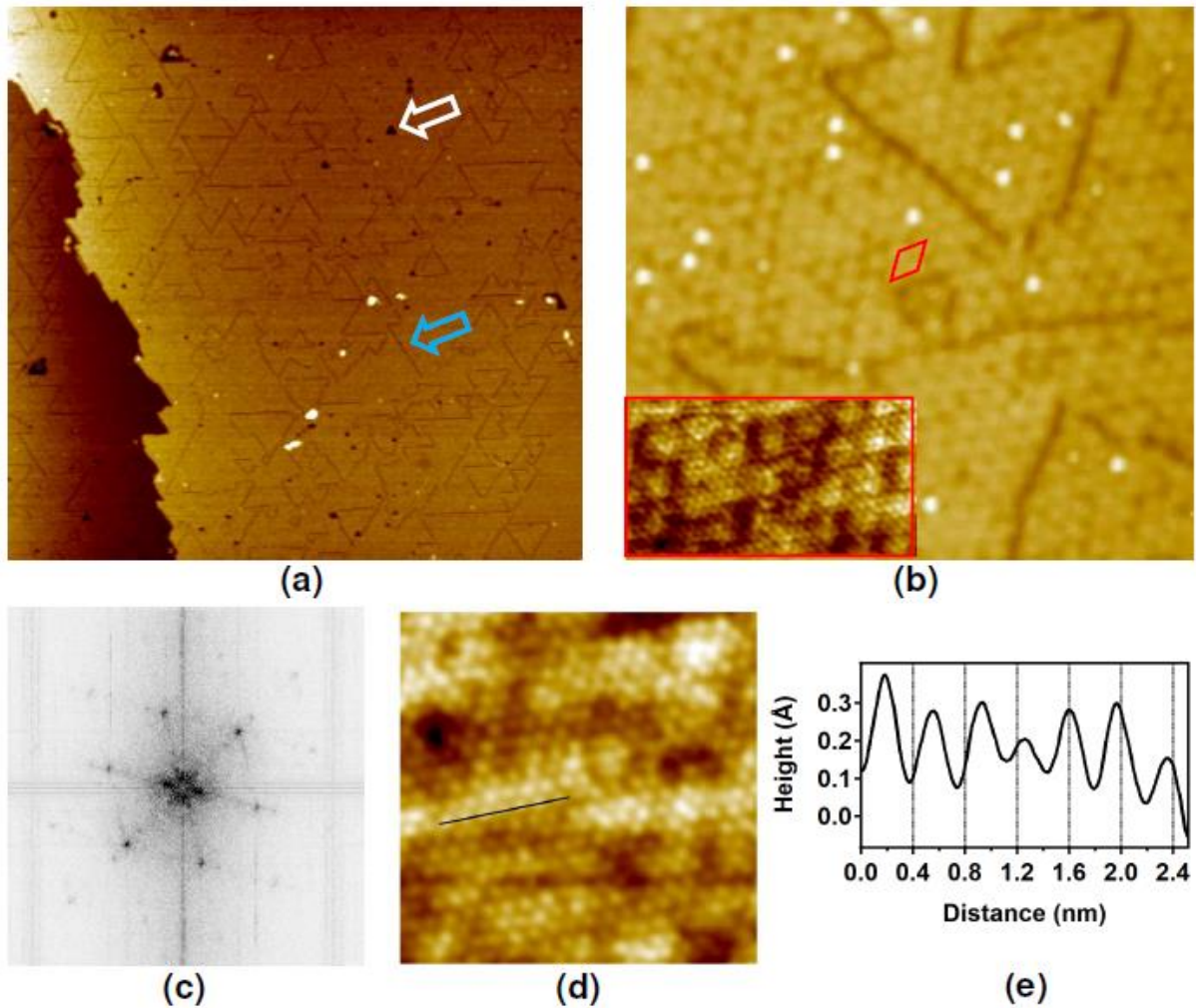

**Figure 2 | Scanning tunneling microscopy. a,** Filled state (-1 V) STM image, $254 \times 254$ nm$^2$. This large area image shows that the arsenene layer forms uniformly across the surface. The white and blue arrows point at two types of defects, i.e., triangular vacancies and straight lines, respectively. **b,** Empty state (2 V) STM image, $31 \times 31$ nm$^2$ revealing features ordered in a $4 \times 4$ periodicity as indicated by the red unit cell. The $4 \times 4$ features appear much weaker at low bias as shown by the STM image in the inset (10 meV, $5.6 \times 10.1$ nm$^2$). This image shows a hexagonal structure consistent with buckled arsenene as well as the Moiré character of the $4 \times 4$ periodicity due to the difference in lattice constants between Ag(111) and arsenene. **c,** Representative fast Fourier transform of STM images showing $4 \times 4$ features as in **b**. **d,** Atomically resolved empty state (10 mV) STM image, $7.5 \times 7.5$ nm$^2$. **e,** Line profile along the

black line in **d** resulting in an average lattice constant of 3.6 Å, which is consistent with buckled arsenene.

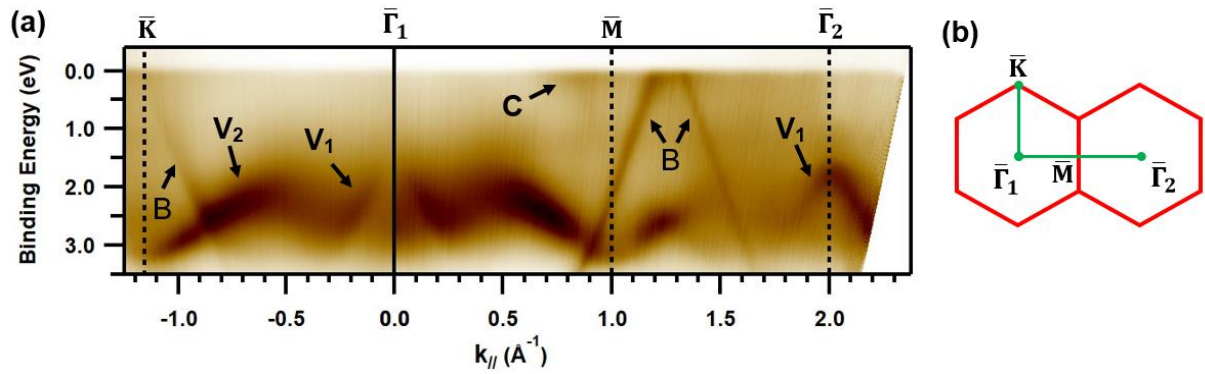

**Figure 3 | Experimental electronic band structure. a,** Band dispersions obtained by ARPES along $\overline{\text{K}}\overline{\Gamma}_1\overline{\text{M}}\overline{\Gamma}_2$ of the $1 \times 1$ SBZ of arsenene using a photon energy of 26 eV. $V_1$ and $V_2$ are valence band dispersions while C originates from the conduction band minimum. B indicates emission from direct transitions between Ag bulk bands. **b,** SBZs with symmetry labels.

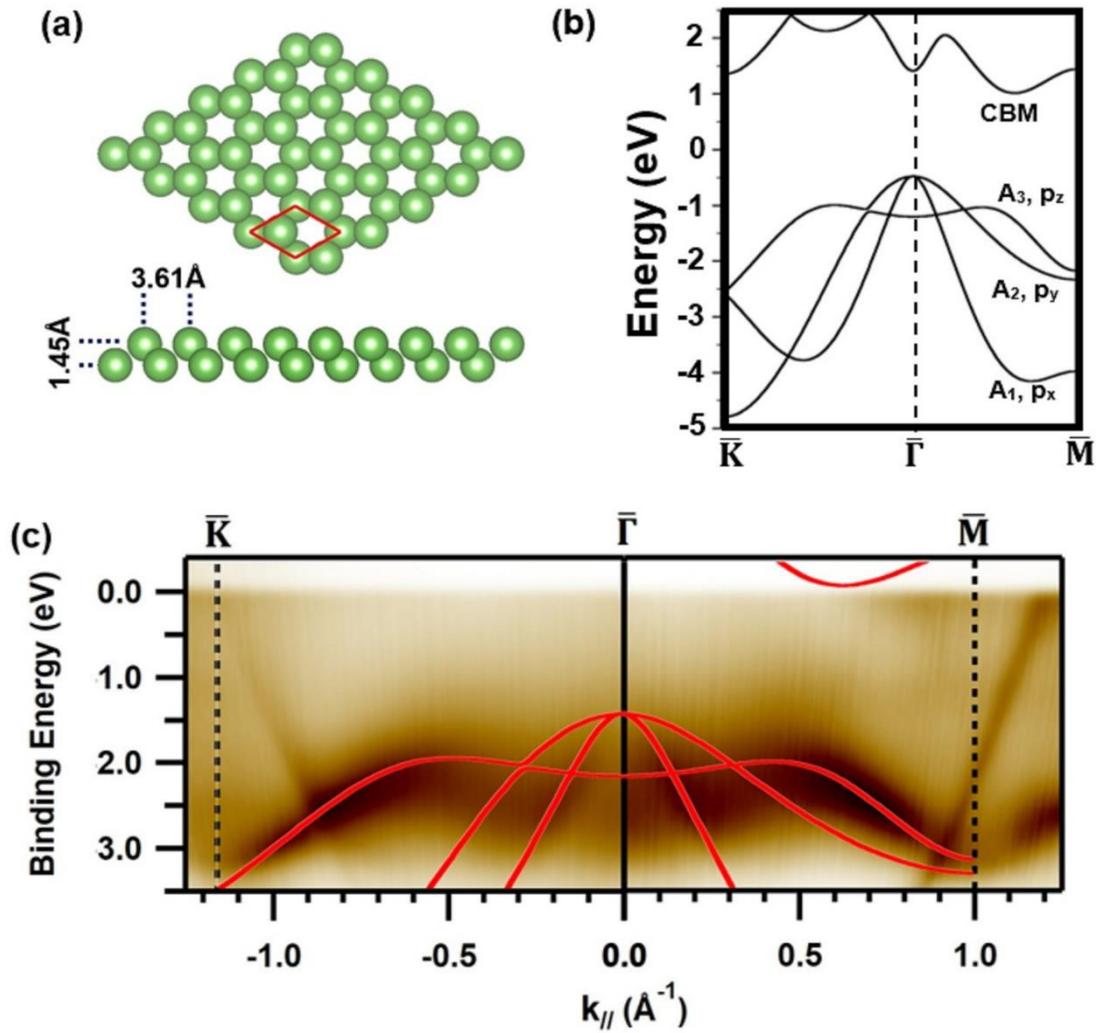

**Figure 4 | Atomic model and calculated electronic band structure of free standing arsenene. a,** Top and side views of the free-standing arsenene model. **b,** Calculated band structure of freestanding arsenene. **c,** Comparison between measured and calculated band structures of arsenene.

15